# Tuning up Fuzzy Inference Systems by using optimization algorithms for the classification of solar flares

## Sintonización de Sistemas de Inferencia Difusa mediante algoritmos de optimización para Clasificación de Fulguraciones Solares

Liz Angélica Ramos Medina[1,3,5], Alex Francisco Bustos Pinzón[1,4,5],
Miguel A. Melgarejo R.[1,5], Santiago Vargas Domínguez[2,5]

1. Universidad Distrital Francisco José de Caldas, 2. Universidad Nacional de Colombia
3. laramosm@correo.udistrital.edu.co, 4. afbustosp@correo.udistrital.edu.co
5. Bogotá D.C., Colombia.

*Abstract*—In this work we describe the implementation and analysis of different optimization algorithms used for finding the best set of parameters for a Fuzzy Inference System intended to classify solar flares. The parameters will be identified among a universe of possible solutions for the algorithms, and the system will be tested in the particular case of dealing with the aim of classifying the solar flares.

*Keywords*—ANFIS, EBDF, fuzzy sets, solar flares.

## 1. INTRODUCTION

The Sun is the main responsible for the varying conditions of the interplanetary medium, particularly, in the space surrounding our planet, in what is commonly known as space weather. Multiple solar phenomena show up at many spatial and temporal scales, and are studied through observations, theoretical models and simulations. Among the most energetic phenomena in the solar system are the solar flares. These are transient events associated to the activity of the star in which certain regions of the solar atmosphere can emit a vast amount of energy up to $10^{25}$ Joules. These zones in the solar atmosphere are associated with the presence of dark spots in the solar surface (photosphere) called sunspots.

Sunspots are the manifestation of intense magnetic fields emerging from the solar interior and crossing the photosphere, inhibiting the normal convection of solar plasma and thus reducing the radiation emission. For this reason the temperature values in sunspots drops approximately 2000 K compared to the temperature in the non-active photosphere, known as quiet sun. Sunspots are proxies of solar activity and their number on the solar disk was used to discover the solar cycle in 1843 [1] and are the main constituents of the so-called solar active regions.

Solar activity has become a very important research topic due to its connection with space weather and the possible impact of energetic phenomena on the normal development of the current technological society, based on satellites, which could be affected by intense solar emissions [2].

Depending on the amount of energy released (flux in $Wm^{-2}$) during the intensity peak of flaring events, solar flares are classified in A, B, C, M or X, as listed in Table 1. The effect of the different types of flares is also different depending on the flare type [3].

Table 1. Classification of Solar Flares. Source: Based on [3].

| Flare class | Peak Flux Range / $Wm^{-2}$ |
|---|---|
| A | $< 10^{-7}$ |
| B | $10^{-7}$ to $10^{-6}$ |
| C | $10^{-6}$ to $10^{-5}$ |
| M | $10^{-5}$ to $10^{-4}$ |
| X | $> 10^{-4}$ |

The main goal of this work is to choose the best Fuzzy Inference System (FIS), from among several FIS tuning methods used, through a validation index Starting from the solar flares characteristics and quantity of them in the solar disk (as inputs of the FIS), each FIS allows to obtain a classification of the solar flares (as output of the FIS). The parameters of each system were tuned using five methods: Manual Tuning, Adaptive Neuro-Fuzzy Inference System (ANFIS) with random initialization [4], Compact Genetic Algorithm (CGA) [5], Differential Evolution (DE) [6] and Stochastic Hill Climbing (SHC) with random initialization [7].

The flow chart that describes the problem is shown in Figure 1, in which the "Problem in Nature" is the unknown way that makes the input values to be related with the output values, observed from Sun behavior. This behavior should be emulated by the FIS. The validation index is a function of the expected output, generated by the Problem in Nature, and from the output obtained by the FIS.

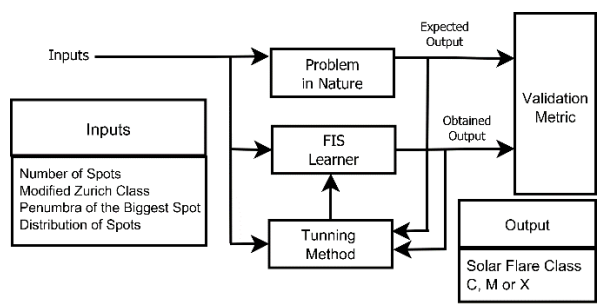

Figure 1. Flow chart of the Global description for the Artificial Intelligence problem. Source: Authors.

The sunspot features and their associated flares were obtained by generating a database according to [2], through a cross search in the sunspots and solar flares catalogs from the National Geophysical Data Center (NGDC). The parameters for the cross search allowed to obtain a total of 1391 individual values, using a time span of 6 hours, in the records from 1999 to 2002, to cover the activity peak of the Solar Cycle 23. The quantities for each class with these parameters are recorded in the Table 2. Note that the generated data presents an imbalance: the number of type C (common) flares are big compared to the M (moderate) flares, data class. Similarly, the M class has more data than X (extreme) flares, as expected from displaying activity of the Sun during its cycle of approximately 11 years.

Table 2. Data used by class. Source: Authors.

| Flare Class | Quantity |
|---|---|
| C | 1194 |
| M | 179 |
| X | 18 |

Aiming to abbreviate, the inputs of the database were numerated as follows:

1. Modified Zurich Class
2. Penumbra: Largest Spot
3. Sunspot Distribution
4. Normalized number of Sunspots

Creating scatter plots from pairs of inputs like in Figure 2, shows that it is not possible to plot a linear function that separates the classes. Also, it is quite clear from the Figure 2 that class M seems to be "absorbed" by class C. Furthermore, class X, having the lower amount of data, is almost not recognizable from class M. Thereby, the attention is focused on classify the class X solar flares.

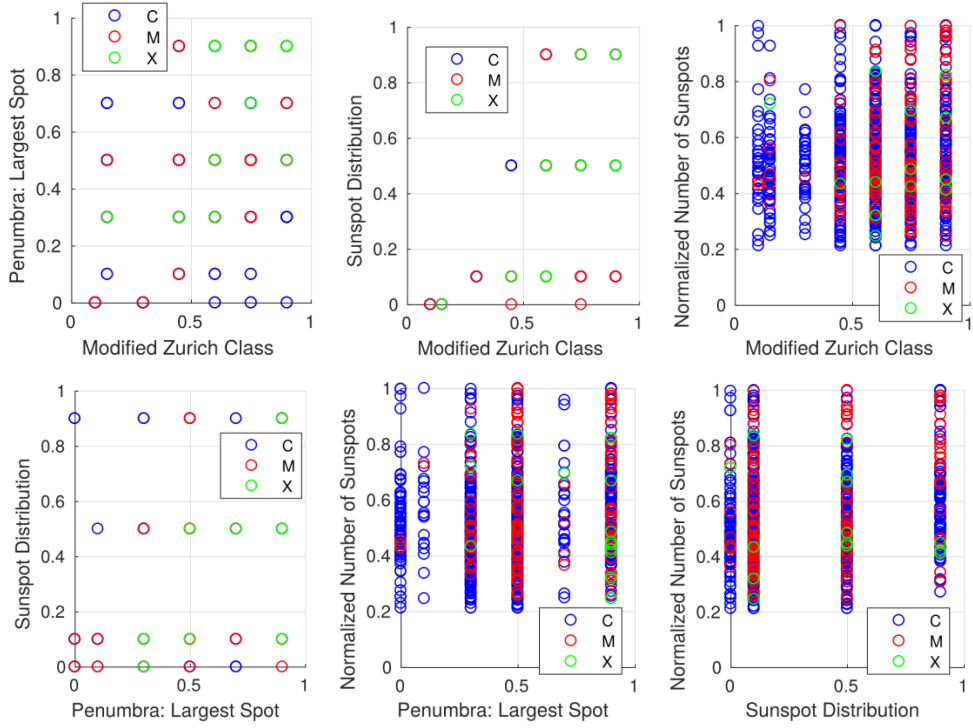
Figure 2. Scatter plots of possible combinations from pairs of the inputs. Source: Authors

## 2. METHODOLOGICAL CONSIDERATIONS

### *2.1 Fuzzy Inference System*

A FIS consists of five components: a base of fuzzy rules, a data base that defines the membership functions of the fuzzy sets used in fuzzy rules, the fuzzy inference engine, the fuzzifier and defuzzifier [4]. The FIS can be represented with an fuzzy basis function expansion in which an input vector $x$ is related with a punctual $y$ output, such that $y = f(x)$. Thus, it is possible to represent in a compact manner the inference process of a FIS and the resulting function is a universal estimator [5].

$$f(x) = \frac{\sum_{l=1}^{M} y_l \prod_{i=1}^{N} \mu_{A_i^l}(x_i)}{\sum_{l=1}^{M} \prod_{i=1}^{N} \mu_{A_i^l}(x_i)} \qquad (1)$$

The FIS represented by (1) has the following characteristics:

- Fuzzification: Singleton
- Membership Functions: Gaussian.
- Implication: Product
- Defuzzification: Average of centers.

The $l$ index refers to the $l$-th rule, being $M$ the total number of rules. By its part, the $i$ index refers to the $i$-th input and $N$ are the total of them. The $\mu_{A_i^l}(x_i)$ membership function (MF) is then unique for each input in every rule. Similarly, the center of the consequent set $y_l$ is unique in every rule [5].

The MFs $\mu_{A_i^l}(x_i)$ are of Gaussian type, and can be written as (2).

$$\mu_{A_i^l}(x_i) = e^{\left[-\frac{(x_i - c_i^l)^2}{2(\sigma_i^l)^2}\right]} \tag{2}$$

Every MF in (2) has their $c$ mean value and a $\sigma$ standard deviation.

The total quantity of parameters that defines a FIS in the form (1) are given by (3), having in mind that, for each input and every rule there are two parameters due to the antecedent set ($c$ and $\sigma$), and an additional parameter being the center of the consequent.

$$C_{Tp} = (2 * M * N) + M \tag{3}$$

## 2.2 Manual Tuning Method

Starting from the authors perceptions about the data and the possible relations that may be present in it, it is possible to create an initial FIS with their fuzzy sets for each of the inputs, their punctual output values, and the rule base allowing to link the fuzzy sets if the inputs to the punctual outputs. The purpose of this method is to deepen into the problem recognizing possible relationships among features as well as revealing preliminary classification rules. Although a valid solution can be found, the most important result of this method is the knowledge derived from approaching the problem.

Initially the software used was GNU's Octave, loading the packages "io" and "fuzzy-logic-toolkit". The first allows that Octave reads the generated CSV dataset, and the second to design, test and verify the manual tuned FIS.

Despite the fact that in the following algorithms the software used was MATLAB, the final FIS created with Octave was migrated to MATLAB through the Fuzzy Logic Designer, a graphical tool part of the Fuzzy Logic Toolbox; with the mere purpose to use the same software tool at the final validation stage.

## 2.3 Adaptive Neuro-Fuzzy Inference System (ANFIS) with random initialization

ANFIS, a FIS based on adaptive networks, is a method based on a supervised learning model that, given a set of input/output pairs $(x, y)$, related by an unknown function $f$, there is an apprentice and a supervisor of the learning process from $f$, with the use of a validation metric to evaluate the results of the apprentice and able to correct it. The algorithm uses a hybrid model that combines least squares method and the decreasing gradient or *back-propagation* method.

In this case the apprentice is a fuzzy system that can be written as the expansion of fuzzy based functions for a Sugeno type system shown in (1). The parameters to be determined correspond to $y_l$, $x_l^i$ and $\sigma_l^i$ [4]. The validation metrics represents the root mean square error (RMSE) between the output value for the fuzzy apprentice system and the output value $y$ of the data pairs [5]. The process aims at minimizing the error for the input values in a set comprising part of the complete available data, which is generally about 70% of them. Searching for an apprentice generalization, it is validated with the remaining 30% of the database.

Additional to the individual (apprentice system with its parameters and rules) to be adjusted, ANFIS requires initial conditions such as the number of rules, number of inputs and the rate of initial learning. For the case mentioned above, the inputs stay constant and the other two parameters are tuned up. Because ANFIS fits the parameters of an existing individual, thus implying a local search, it executes several times and, prior to this, it generates the individual with initialized parameters in random values, aiming at (depending on randomization) perform a global search in a whole universe of possible solutions.

Algorithm 1. Pseudo code for the MATLAB implementation using the ANFIS function. (Source: Authors)

```
1:   Training = 70% of Base
2:   Validation = 30% of Base
3:   Vector of rules to be tested R_T
4:   Vector of Initial Learning Rates to be tested TA_T
5:   n = number of tests
6:   Ep = number of epochs
7:   for i ∈ TA_T do
8:     for j ∈ R_T do
9:       for l = 1: n do
10:         Generate random FIS with 4 inputs and R = j
11:         Evaluate ANFIS function with TA = i, Ep, Training, Validation and random generated FIS
12:         Save the FIS with lowest validation error, the training error and the output vector validation k.
13:       end for
14:     end for
15:   end for
16:   Lowest validation error = MinV, associated FIS = CheckFis
```

## 2.4 Compact Genetic Algorithm (CGA)

This belongs to a series of algorithms known as Probabilistic Model Building Genetic Algorithm (PMBGA) [8], which are characterized by discriminating the significant contribution attributes in the construction of an optimal individual. The validation indexes for determining the performance of an individual is the "Fitness" function, which in turn depends on the problem to be solved. The implementation considers an individual with the best performance when the value of this function is minimized.

Because in this work we are dealing with a classification problem, besides using the RMSE, we decided to also consider the use of classification error and correlation. With that in mind, we can assemble an initial brief of a fitness function (4).

$$F = (E_{CC} + E_{CM} + E_{CX})^2 \times E_{Rmse} \times (1 - \rho)^2 \qquad (4)$$

And

$$E_{Cx} = \frac{h_x}{C_x} w_x \qquad (5)$$

Where:

$E_{Cx}$ : Classification error for the class $x$
$h_x$ : Number of bad classified data for class $x$
$C_x$ : Total number of data for class $x$
$w_x$ : Weight assigned to the classification error of class $x$
$E_{Rmse}$ : Root Mean Square Error
$\rho$ : Correlation

Every $E_{Cx}$ classification error has its respective $w_x$ weight. As the database is inherently imbalanced, every weight $w_x$ was assigned to be greater than the proportion of data belonging to class C, to the quantity of data from the other classes:

$$w_M > \frac{1194}{179} \sim 6.7 \rightarrow w_M = 10$$

$$w_X > \frac{1194}{18} \sim 66 \rightarrow w_X = 100$$

Therefore, the weight associated to the class X of solar flares, for which the number of data is lower, has the highest value. By doing this, a badly classified data that belongs to this class produces a more significant increase in the first factor of (4) that one not incorrectly classified in class C, in the final fitness function factors (6)

$$E_{CC} = \frac{h_C}{C_C}, E_{CM} = \frac{h_M}{C_M} \times 10, E_{CX} = \frac{h_X}{C_X} \times 100 \tag{6}$$

To explain the $E_{Rmse}$ Root Mean Square Error in (4), suppose that the problem is not a classification problem, but a prediction problem instead. For a conceptual brief, the $E_{Rmse}$ gives an idea on how the individual are *not* "following" the expected sequence from the training data [5]. Then, a bad predictor will have a greater $E_{Rmse}$ value, than other that gets closer to the output values of the database, and considering that the data also depends on some time unit. The root mean square error is mathematically described as:

$$RMSE = \sqrt{\frac{1}{n}\sum_{i=1}^{n} e_i^2} \tag{7}$$

$$e = (v_o - v_e) \tag{8}$$

Where

- $v_o$ is the value obtained
- $v_e$ is the expected value

The number of rules was taken from the obtained result with the ANFIS algorithm, $R = 8$ rules. For developing the algorithm, the parameter for adjusting the converging speed of the probability vector $n$ is tuned. Since the optimal value is unknown, it is randomly designated based on [5], and implemented in MATLAB. The process of randomly varying $n$ and developing the algorithm, is repeated several times ($w$ = number of experiments). Finally, among the best solutions the value generating the lowest number in (4) with (6) is found.

The parameters describing every FIS (individual) are then converted from real to binary data, due to the method adjusting every bit.

---

Algorithm 2. Pseudo code for CGA (Based on [5])

1: Training = 70% of Base
2: Validation = 30% of Base
3: $w$ = number of tests
4: $N_p$ = number of parameters
5: $b_p$ = number of bits per parameter
6: $a = N_p * b_p$
7: $n$ = probability adjustment parameter
8: $p$ = probability vector
9: $N_i$ = number of individuals
10: $I_{alea}$ = vector of $N_i$ individuals
11: **for** $i = 1\ to\ w$ **do**
12:    $n$ = Random value
13:    **for** $l = 1: N_i$ **do**
14:       $I_{alea}(l)$ = Random FIS
15:       Evaluate and order individuals so that the best is in position $I_{alea}(1)$
16:    **end for**
17:    **for** j= 2: $N_i$ **do**
18:       Winner, Loser = competition $(I_{alea}(1), I_{alea}(j))$
19:       **for** g = 1: $a$ **do**
20:          **if** Winner(g) ~ Loser(g) **then**
21:             **if** Winner(g) = 1 **then** $p(g) = p(g) + 1/n$
22:             **else** $p(g) = p(g) - 1/n$
23:          **end if**

```
24:         end if
25:       end for
26:    end for
27:    for g = 1: a do
28:       if p(g) > 0 and p(i) < 1 then
29:          go to step 13
30:       end if
31:    end for
32: end for
```

## 2.5 Differential Evolution

This is an algorithm based on the evolution of a population of vectors (individuals) with real parameters, which represent solutions in the searching space.

The algorithm of differential evolution is basically composed by 4 steps, as follows:

- Initialization: Every vector (individual) of the population is randomly initialized.
- Mutation: A mutation is applied in order to create a testing population of individual.
- Crossing: Every vector is used as a mutant vector.
- Selection: The testing vector previously obtained is used to do the crossing procedure, which compete with the target vector by the evaluation of the Fitness function. [6]

Algorithm 3. Pseudo code for DE (Source: Based on [6]).

```
1:  Training = 70% of Base
2:  Validation = 30% of Base
3:  f_m = mutation constant
4:  c_r = crossover constant
5:  N_i = number of individuals
6:  N_g = number of generations
7:  w = number of tests
8:  V_i = individuals vector
9:  N_p = number of parameters
10: V_o = target vector
11: V_m = mutation vector
12: V_c = crossover vector
13: b_i = vector of the best individual
14: for i = 1 to w do
15:    for l = 1: N_i do
16:       V_i(l) = Random FIS
17:       Evaluate individuals with the fitness function (4)
18:    end for
19:    for j= 1: N_g do
20:       for g = 1: N_i do
21:          V_o = V_i(g)
22:          Sort the individuals from best to worst according to (4)
23:          b_i = Vi(1)
24:          V_m = mutation(b_i, f_m)
25:          for k = 1: N_p do
26:             V_c = cross(V_o, V_m, c_r)
27:          end for
28:          if V_c is better than V_o then
29:             replace V_o with V_c
30:          else keep V_o
31:          end if
32:       end for
33:    end for
34: end for
```

## 2.6 Stochastic Hill Climbing (SHC) with random initialization

The Stochastic Hill Climbing, consist on taking a FIS (1) and keep evaluating the solutions in the vicinity of it [7, 9] in a maximum number of iterations. The parameters of the input FIS are randomly initialized.

Algorithm 4. Pseudo code for Stochastic Hill Climbing. [10]

1: **Require:** $I_{max}$, Dimensions
2: **Ensure:** $Current$
3:   $Current \leftarrow$ RandomSolution(Dimensions)
4: **for** $iter_i \in I_{max}$ **do**
5:     $Candidate \leftarrow$ RandomNeighbor($Current$)
6:     **if** Cost($Candidate$) ≤ Cost($Current$) **then**
7:       $Current \leftarrow Candidate$
14:     **end if**
15: **end for**

Where:

$I_{max}$ : Maximum number of iterations
$Sol$ : Some particular solution (like $Current$ or $Candidate$)
Cost($Sol$) : Fitness function, obeys (2)

RandomNeighbor($Current$) need also the center and deviation variations, that refers to the allowed absolute value variations of the related parameters when searching for a neighbor. As example, if some of the parameters has the value 0.6, and the specified variation of this parameter is 0.1, then the neighbor will have some uniformly distributed random value between 0.5 and 0.7.

Every separate experiment consist on a single run of a program that implements the algorithm 4, to obtain a final single individual, but $n$ individuals can be obtained by running $n$ experiments. Afterwards, the individuals can be evaluated with (4) and the validation base, in order to choose the best individual of the $n$ individuals.

## 2.7 Confusion Matrixes

The classifier output consists on C values, corresponding to the $\omega_1, \omega_2, \ldots, \omega_c$ classes. Due to the erroneous classifications occasionally occurring, the multiclass sorter is evaluated through a $(C \times C)$ – dimensional confusion rate matrix showing the respective classification errors between classes (off diagonal) and correct classifications (diagonal elements). [11]

Table 3. Confusion Matrix for a three class sorting problem. Source: Based on [11].

|  |  | Predicted Class | | |
|---|---|---|---|---|
|  |  | $\omega_1$ | $\omega_2$ | $\omega_3$ |
| **Actual Class** | $\omega_1$ | $C_{\omega_{1,1}}$ | $C_{\omega_{1,2}}$ | $C_{\omega_{1,3}}$ |
|  | $\omega_2$ | $C_{\omega_{2,1}}$ | $C_{\omega_{2,2}}$ | $C_{\omega_{2,3}}$ |
|  | $\omega_3$ | $C_{\omega_{3,1}}$ | $C_{\omega_{3,2}}$ | $C_{\omega_{3,3}}$ |

Table 3 shows an example of a confusion matrix for a total of $C = 3$ classes. The $C_{\omega_{i,j}}$ elements correspond to the data quantity from the $\omega_i$ class that was classified as elements of the $\omega_j$ class.

## 3. PARAMETERS FOR THE ALGORITHMS

Excluding the manual tuned FIS, and in order to allow the replicability of similar results, we expose briefly the parameters used for the algorithms. For the CGA, DE and SHC algorithms, the number of rules was taken from the best ANFIS result, as shown in Table 3.

### 3.1 Manual Tuning

With this method, the FIS finally had the characteristics recorded in Table 4.

Table 4. Classification of Solar Flares. Source: Authors.

| Parameter | Value |
|---|---|
| MFs for Input 1 | 7 |
| MFs for Input 2 | 6 |
| MFs for Input 3 | 4 |
| MFs for Input 4 | 3 |
| MFs for the Output | 3 |
| Rules | 8 |

As the parameters for this method obey to human perceptions of the problem, only the main features are shown in Table 4, for this reason this method was applied only as an exercise of comparison between the human performance and machine performance, in building a FIS that solves the classification problem. These values are not normative by the same fact that the parameters were based from human perceptions of the authors, are then allowed to test other values, but the manual tuning method takes too much time to get a single FIS.

### 3.2 Adaptive Neuro-Fuzzy Inference System (ANFIS) with random initialization

Table 5. Initialization Parameters for the implementation of ANFIS with random initialization. Source: Authors.

| Parameter | Value |
|---|---|
| Epochs | 500 |
| Number of experiments | 200 |
| Tested learning rates (TA) | 0.01, 0.1, 1 |
| Tested number of rules | 8, 14, 16, 32 |
| Performance Function | Root Mean Square Error (7) |

### 3.3 Compact Genetic Algorithm (CGA)

Table 6. Initialization Parameters for the CGA implementation. Source: Authors.

| Parameter | Value |
|---|---|
| Number of rules | 8 |
| Number of parameters to optimize | 72 |
| Number of bits per parameter | 8 |
| Binary coding method | Sign-magnitude |
| Population size | 30 |
| Number of experiments | 500 |
| Maximum number of generations | 10000 |
| Stop criterion | Convergence of probability vector and error repetition |
| Performance Function | Fitness function (4) |

## 3.4 Differential Evolution (DE)

Table 7. Initialization Parameters for the DE Algorithm implementation. Source: Authors.

| Parameter | Value |
|---|---|
| Number of rules | 8 |
| Number of parameters to optimize | 72 |
| Population size | 30 |
| Number of generations | 50 |
| Mutation constant | 0.5 |
| Crossover constant | 0.9 |
| Number of experiments | 500 |
| Variant | ED/best/1/bin |
| Stop criterion | Number of generations and number of experiments |
| Performance Function | Fitness function (4) |

## 3.5 Stochastic Hill Climbing (SHC) with random initialization

Once the base individual for the Hill Climbing was randomly initialized, the SHC algorithm used the parameters listed in Table 8.

Table 8. Parameters for the implementation of the SHC with random initialization algorithm. Source: Authors.

| Parameter | Value |
|---|---|
| Number of rules | 8 |
| Number of parameters to optimize | 72 |
| Number of experiments | 10 |
| Number of iterations by experiment | 8000 |
| Center variation | 0.1 |
| Deviation variation | 0.5 |
| Stop criterion | Number of iterations |
| Performance Function | Fitness function (4) |

## 4. RESULTS

In this section are firstly shown the best results for every method and their analysis. This analysis includes a comparison of their performance.

## 4.1 Confusion Matrices

The best FIS obtained by each algorithm was evaluated using the whole database. With the evaluated output values and the expected output values a confusion matrix can be filled as shown in Table 3 to obtain the matrices shown in Tables 9, 11, 12, 13 and 14.

Table 9. Confusion Matrix for the manual tuned FIS. Source: Authors.

|  |  | Predicted Class | | |
|---|---|---|---|---|
|  |  | C | M | X |
| **Actual Class** | C | 0 | 815 | 198 |
|  | M | 0 | 100 | 44 |
|  | X | 0 | 12 | 6 |

In the case of ANFIS, the individual with the lowest validation error was selected for each of the different combinations of number of rules and initial learning rate (LR) as shown in Table 10.

Table 10. List of the lowest validation error (RMSE) for every $n$ test. Source: Authors

| LR | Number of Rules | | | |
|---|---|---|---|---|
| | 8 | 14 | 16 | 32 |
| 0,01 | 0.3681 | 0.3688 | 0.3708 | 0.3747 |
| 0,1 | 0.3667 | 0.3682 | 0.3705 | 0.3735 |
| 1 | 0.3658 | 0.3661 | 0.3667 | 0.3687 |

From Table 10 the best individual are chosen to make the confusion matrix shown in Table 11. In order to compare the results with the same metric, this individual was evaluated with (4) and its results are part of Table 15. The chosen individual was obtained with the following parameters:

- $Rules = 8$
- $Learning\ Rate\ (LR) = 1$

Table 11. Confusion Matrix for the best ANFIS individual chosen. Source: Authors.

| | | Predicted Class | | |
|---|---|---|---|---|
| | | $C$ | $M$ | $X$ |
| Actual Class | $C$ | 1133 | 0 | 0 |
| | $M$ | 155 | 0 | 0 |
| | $X$ | 15 | 0 | 0 |

The best FIS obtained by the CGA occurred on experiment $w = 175$ and for a value $n = 41$ of the probability adjustment parameter.

Table 12. Confusion Matrix for the best individual obtained by CGA. Source: Authors.

| | | Predicted Class | | |
|---|---|---|---|---|
| | | $C$ | $M$ | $X$ |
| Actual Class | $C$ | 1194 | 0 | 0 |
| | $M$ | 179 | 0 | 0 |
| | $X$ | 18 | 0 | 0 |

**Table 13. Confusion Matrix for the best individual obtained by the DE Algorithm. Source: Authors.**

<table>
<tr><td colspan="2" rowspan="2"></td><th colspan="3">Predicted Class</th></tr>
<tr><th>C</th><th>M</th><th>X</th></tr>
<tr><th rowspan="3">Actual Class</th><th>C</th><td>0</td><td>552</td><td>552</td></tr>
<tr><th>M</th><td>0</td><td>57</td><td>113</td></tr>
<tr><th>X</th><td>0</td><td>0</td><td>18</td></tr>
</table>

**Table 14. Confusion Matrix for the best individual obtained by the SHC Algorithm. Source: Authors.**

<table>
<tr><td colspan="2" rowspan="2"></td><th colspan="3">Predicted Class</th></tr>
<tr><th>C</th><th>M</th><th>X</th></tr>
<tr><th rowspan="3">Actual Class</th><th>C</th><td>0</td><td>0</td><td>1194</td></tr>
<tr><th>M</th><td>0</td><td>0</td><td>179</td></tr>
<tr><th>X</th><td>0</td><td>0</td><td>18</td></tr>
</table>

## 4.2 Final Results by the Validation Metric

Table 15 lists the more relevant metrics for the individuals in every scheme. The final individual was the one with the lowest value of the Fitness function (4), using the validation database.

**Table 15. Validation Errors for the best functions obtained. Source: Authors.**

| Method | Fitness | $E_{CC}$ | $E_{CM}$ | $E_{CX}$ |
|---|---|---|---|---|
| Manual | 13282 | 1 | 3,7736 | 61,111 |
| ANFIS | 4487.2 | 0.0446 | 10 | 100 |
| CGA | 3895.1 | 0 | 10 | 100 |
| DE | 33.2454 | 1 | 4.9057 | 0 |
| SHC | 96.712 | 1 | 10 | 0 |

As shown in Table 15, the individual with the lowest value for the fitness function was obtained was the DE algorithm, since it was able to classify all samples from class X and additionally 57 from M class (see Table 13). Using SHC with random initialization method it was possible to classify only samples of class X. For the ANFIS with random initialization and CGA algorithms it was not possible to classify any of the samples of class X or M.

## 4.3 Statistical Analysis

To perform a statistical analysis of the algorithms implemented, the Welch's t-test was used for two-samples, assuming unequal variances to confirm or reject the null hypothesis whether both methods provide similar analytical results or not. [12]

**Table 16. Results for the Welch's t-test between DE and CGA. Source: Authors.**

|  | DE | CGA |
|---|---|---|
| Mean | 2074.216694 | 15808.9081 |
| Variance | 30477597.49 | 45636475.15 |
| Observations | 500 | 500 |
| Hypothetical difference of means | 0 | |
| Degrees of freedom | 960 | |
| Statistic t | -35.20233147 | |
| P (T ≤ t) one tail | 2.7328E-175 | |
| Critical value of t (one tail) | 1.646442429 | |
| P (T ≤ t) two tails | 5.4656E-175 | |
| Critical value of t (two tails) | 1.962438166 | |

**Table 17. Results for the Welch's t-test between DE and ANFIS with Random Initialization. Source: Authors.**

|  | DE | ANFIS with Random Initialization |
|---|---|---|
| Mean | 2074.216694 | 5737.628906 |
| Variance | 30477597.49 | 184874.1979 |
| Observations | 500 | 2400 |
| Hypothetical difference of means | 0 | |
| Degrees of freedom | 500 | |
| Statistic t | -14.82880602 | |
| P (T ≤ t) one tail | 8.5552E-42 | |
| Critical value of t (one tail) | 1.647906854 | |
| P (T ≤ t) two tails | 1.71104E-41 | |
| Critical value of t (two tails) | 1.964719837 | |

**Table 18. Results for the Welch's t-test between DE and SHC. Source: Authors.**

|  | DE | SHC |
|---|---|---|
| Mean | 2074.216694 | 2001.247664 |
| Variance | 30477597.49 | 19996528.91 |
| Observations | 500 | 500 |
| Hypothetical difference of means | 0 | |
| Degrees of freedom | 957 | |
| Statistic t | 0.229662015 | |
| P (T ≤ t) one tail | 0.409201745 | |
| Critical value of t (one tail) | 1.646447414 | |
| P (T ≤ t) two tails | 0.818403489 | |
| Critical value of t (two tails) | 1.962445932 | |

Comparing the results of the test between ED with the CGA and ANFIS algorithms as shown in Tables 16 and 17 respectively, it is possible to reject the null hypothesis and conclude that the methods provide different analytical results with a 99% confidence level.

On the other hand, from Table 18 it can be evidenced that, although the best solution was achieved with the DE algorithm, the average and the variance of the fitness of the individuals obtained with SHC are better than those obtained with DE. This result makes sense in the light of the non free lunch theorems [13], which state that optimization methods perform similarly in average over the entire set of possible optimization problems. The result of the Welch's t-test shows that the null hypothesis should not be rejected because in the case of two tails the confidence level to reject is less than 20% and in the case of one tail it is less than 60%. Therefore, both methods provide the same average results and the observed differences are purely due to random errors.

# 5. CONCLUSIONS

In this section we summarize the obtained results and discuss on the different aspects of their performance.

- Due to the imbalance in the database, systems and algorithms used in the present work have limited options to learn from class M, and much lower ones from class X.
- Additionally for ANFIS, because of the fact mentioned before, the validation metrics for RMSE is not adequate for solving the problem since it ignores the classification error, from which it is evidenced that the best individual obtained in this method is an optimal class C classifier, but not so for the rest of classes.
- Despite the Compact Genetic Algorithm has a simple description with little memory, it sufficiently restricts the space of solutions since it works with parameters represented in fixed point, having a more reduced universe as compared to the representation in floating points.

From the items listed above, and from Table 4, it cannot be discarded different problems in which either class C are distinguished from being or not solar flares (modifying the generation parameters of the database), or type M or X solar flares are distinguished. As a future work, the problem can be addressed by using neural network algorithms, e.g. *Cascade-Correlation Neural Networks* (CCNNs), *Support Vector Machines* (SVMs) and *Radial Basis Function Networks* (RBFNs) [2] instead of FISs, in order to determine if it is feasible to obtain a best classifier and therefore extend the problem of estimating the occurrence of solar flares.